\documentclass[conference]{IEEEtran}
\usepackage{amsmath}
\usepackage{latexsym}
\usepackage{amssymb}
\usepackage{amsfonts}
\usepackage{graphicx}
\usepackage{bbding}
\usepackage{indentfirst}
\usepackage{cases}
\usepackage{algorithm,algorithmic}
\usepackage{subeqnarray}
\usepackage{bm}
\usepackage{color}
\IEEEoverridecommandlockouts
\allowdisplaybreaks[4]

\begin{document}
\title{Joint Activity Detection and Channel Estimation for mmW/THz Wideband Massive Access}
\author{Xiaodan Shao, Xiaoming Chen, Caijun Zhong, and Zhaoyang Zhang
\\College of Information Science and Electronic Engineering, Zhejiang University, Hangzhou, China.}\maketitle

\begin{abstract}
Millimeter-wave/Terahertz (mmW/THz) communications have shown great potential for wideband massive access in next-generation cellular internet of things (IoT) networks. To decrease the length of pilot sequences and the computational complexity in wideband massive access, this paper proposes a novel joint activity detection and channel estimation (JADCE) algorithm. Specifically, after formulating JADCE as a problem of recovering a simultaneously sparse-group and low rank matrix according to the characteristics of mmW/THz channel, we prove that jointly imposing $l_0$ norm and low rank on such a matrix can achieve a robust recovery under sufficient conditions, and verify that the number of measurements derived for the mmW/THz wideband massive access system is significantly smaller than currently known measurements bound derived for the conventional simultaneously sparse and low-rank recovery. Furthermore, we propose a multi-rank aware method by exploiting the quotient geometry of product of complex rank-$L_{\max}$ matrices with the maximum number of scattering clusters $L_{\max}$. Theoretical analysis and simulation results confirm the superiority of the proposed algorithm in terms of computational complexity, detection error rate, and channel estimation accuracy.
\end{abstract}


\IEEEpeerreviewmaketitle
\section{Introduction}
The rapid development of Internet-of-Things (IoT) in various fields requires the next-generation cellular IoT networks to support massive connectivity for an exponential growing in the number of machine-type devices \cite{Book}. In this context, grant-free random access protocol, as a key of massive connectivity, has gained considerable attentions in recent years \cite{free}. Specifically, active devices access wireless networks by transmitting pre-assigned pilot sequences without a grant, and then the base station (BS) jointly detects the device activity and estimates channel state information (CSI). Consequently, access latency and signaling overhead can be significantly reduced, especially in the scenario of massive connectivity \cite{massive}.

As joint activity detection and channel estimation (JADCE) is a typical sparse problem due to sporadic IoT applications, some compressed sensing based algorithms have recently been proposed. The authors in \cite{AMPO} and \cite{AMPO1} proposed an approximated message passing (AMP) algorithm that exploited the statistics of the wireless channel for JADCE. Moreover, the authors in \cite{covar1,jia} proposed a low-complexity algorithm for device activity detection, which only requires the covariance matrix of the received signal. A common of previous works is that they all considered a narrowband multiple access system. In fact, to satisfy the requirements of massive connectivity and huge capacity, the next-generation cellular IoT may adopt millimeter-wave or even terahertz wideband communication techniques \cite{mmw}. In the scenario of wideband massive access, there are two challenging issues. Firstly, JADCE over wideband channels requires very long pilot sequences, resulting in large signaling overhead. Secondly, JADCE over wideband channels is a large-dimensional signal processing problem due to the use of a ultra large-scale antenna array at the BS, resulting in prohibitive computational complexity. Thus, it is necessary to design a feasible and effective wideband massive access scheme based on the characteristics of mmW/THz channels. Extensive experiments show that mmW/THz channels spread in the form of clusters of paths in the angular domain, which leads to a structured sparsity pattern that can be exploited to enhance the estimation performance with low complexity and overhead. In addition, mmW/THz channels exhibit a joint sparse and low-rank structure where the rank is far smaller than the sparsity level of the channel. Leveraging the joint sparse and low-rank structure, a channel estimator for a single device was proposed in \cite{single}, where the sparse and low-rank properties were respectively utilized in two consecutive stages, but not in the joint sense. Note
that the low-rank constraint is in general NP-hard, most related works adopted a nuclear norm to relax the rank constraint. However, the nuclear norm based convex relaxation approaches fail to well incorporate the fixed-rank matrices for sparse signal recovery due to the poor structures. As a result, conventional JADCE algorithms can not provide satisfactory detection and estimation performance.

To better exploit the joint sparse and low-rank structure of mmW/THz channels for wideband massive access, this paper designs a multi-rank aware JADCE algorithm for cellular IoT. The contributions of this paper are as follows:

\begin{enumerate}

\item The low-rank and sparse properties in the delay-angular domain of mmW/THz channels are analyzed and exploited to design a novel wideband JADCE algorithm.

\item The sparse-group and low-rank restricted isometry property (SG$\&$L-RIP) of the proposed algorithm is analyzed. Moreover, theoretical analysis proves that the proposed algorithm has low computational complexity.

\item Simulation results show that the proposed algorithm can achieve near-optimal detection and estimation performance.

\end{enumerate}

\section{System Model and Problem Formulation}
We consider a mmW/THz wideband cellular IoT network, where a BS equipped with $M$ antennas serves $N$ single-antenna IoT devices. To effectively exploit the wideband benefits of mmW/THz, the OFDM modulation scheme with $B$ subcarries is adopted. Due to the burst characteristic of IoT applications, only a fraction of IoT devices are active at any given time slot. To reduce access latency, a grant-free random access scheme is usually utilized. In what follows, we first introduce the mmW/THz channel and then formulate a wideband joint activity detection and channel estimation (JADCE) problem.

\subsection{mmW/THz Channel Model in Delay-Angular Representation}
Without loss of generality, we focus on the channel $\mathbf{H}_n\in \mathbb{C}^{M\times B}$ from the BS to the $n$th IoT device over $B$ subcarries. Since the mmW/THz channel $\mathbf{H}_n$ is a superposition of a small number $L_n$ of resolvable propagation paths characterized by their delay/angle pairs. Due to the high resolution in the spatial domain by deploying an ultra-large-scale antenna array at the BS and in the frequency domain by utilizing ultra-wide mmW/THz band, the channel $\mathbf{H}_n$ exhibits angular and delay spreads. Hence, the channel $\mathbf{H}_n$ can be expressed as
\begin{eqnarray}\label{sch}
\mathbf{H}_n\!\!\!\!\!\!&=&\!\!\!\!\!\!\sum_{l=1}^{L_n} \left( \sum_{j=1}^{J_n }\iota_{n,l,j}\mathbf{a}(\theta_{n,l}-\phi_{n,l,j})\right )\nonumber\\
\!\!\!\!\!\!&&\!\!\!\!\!\!*\left( \sum_{i=1}^{I_n}\xi _{n,l,i}\mathbf{b}^H(\tau_{n,l}-\varphi _{n,l,i})\right ),
\end{eqnarray}
where $L_n$ represents the number of resolvable paths, $\phi_{n,l,j}$ denotes the angle shift with respect to the mean angle $\theta_{n,l}$, $\varphi _{n,l,i}$ denotes the delay shift with respect to the mean delay $\tau_{n,l}$, $I_n$ and $J_n$ represent the number of delay shifts and angle shifts, $\xi _{n,l,i}$ denotes the complex gain of the $l$th path with the $i$th delay shift, and $\iota_{n,l,j}$ denotes the complex gain of the $l$th path with the $j$th angle shift. $\theta _{n,l} \in [0, 1]$ and $\tau _{n,l}\in [0, \gamma T_s]$ are the mean angle of arrival and delay of the $l$th path with $\gamma \leq 1$ and $T_s$ being an OFDM symbol duration. Define $\phi\in[-\pi/2, \pi/2]$ and $\theta=d\sin(\phi)$, where $d$ is the normalized antenna spacing. Suppose the BS employs a uniform linear array (ULA) antenna array whose angle and delay response vectors are given by $
\mathbf{a}(\theta)=[1,e^{-j2\pi\theta},\cdots ,e^{-j2\pi(M-1)\theta}]^T,
$ and $
\mathbf{b}(\tau)=[1,e^{-j2\pi\tau/T_s},\cdots,e^{-j2\pi\tau(B-1)\tau/T_s}]^T,
$
respectively \cite{mmchan}. In this paper, we use $(\cdot)^H$ and $(\cdot)^T$ to denote conjugate transpose and transpose respectively. According to the delay-angular domain characteristics, the
mmW/THz channel $\mathbf{H}_n$ can be rewritten as
\begin{eqnarray*}
\mathbf{H}_n
=\mathbf{A}_{\theta}\left(\sum_{l=1}^{L_n}\boldsymbol{\iota}_{n,l}\boldsymbol{\xi}_{n,l}^T\right)\mathbf{A}_{\tau}^H
\triangleq \mathbf{A}_{\theta}\widetilde{\mathbf{X}}_n\mathbf{A}_{\tau}^H,
\end{eqnarray*}
where $\widetilde{\mathbf{X}}_n\in \mathbb{C}^{M \times D}$ is the delay-angle representation channel, $\mathbf{A}_{\theta}$ and $\mathbf{A}_{\tau}$ are sampled versions of the interval [0, 1] and the interval [0, $T_s$] respectively, which can be represented as
$
\mathbf{A}_\theta \triangleq[\mathbf{a}(0),\cdots, \mathbf{a}((M-1)/M)]\in\mathbb{C}^{M\times M}$ and
$\mathbf{A}_\tau \triangleq [\mathbf{b}(0),\mathbf{b}(T_s/B),\cdots,\mathbf{b}((D-1)T_s/B)]\in\mathbb{C}^{B\times D}$,
where $D=\lfloor{\gamma  B}\rfloor$ denotes the channel delay spread in samples.
$\boldsymbol{\xi}_{n,l}\in \mathbb{C}^{D}$ and $\boldsymbol{\iota}_{n,l}\in \mathbb{C}^{M}$ are the virtual representation of delay and angular gains over the $l$th path, respectively.

\begin{figure}[h]
  \centering
\includegraphics [width=0.23\textwidth] {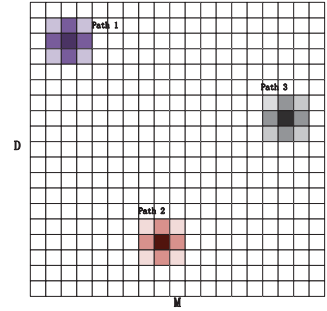}
\caption{Illustration of the sparse structure for delay-angular representation of channel in the
presence of delay and angular spreads, with number of paths $L=3$, $p=3$ and $D=M=19$. Suppose that the effective entries are concentrated in squares in channel matrix painted with colors and the values of white entries are nearly zero.}
\label{channel}
\end{figure}

Note that the delay-angular representation channel $\widetilde{\mathbf{X}}_n$ is a sparse matrix with at most $p_n^2L_n$ nonzero entries out of a total $DM$. Also, $\widetilde{\mathbf{X}}_n$ has at most $p_nL_n$ nonzero rows and $p_nL_n$ nonzero columns, in which $p_nL_n\ll\text{min}(D, M)$ due to the limited scattering nature and small angular spreads of the mmW/THz wideband signal, as shown in Fig. \ref{channel}. In particular, $\widetilde{\mathbf{X}}_n$ has a low-rank structure with $\text{rank}(\widetilde{\mathbf{X}}_n)=L_n$ which is much less than the sparsity of the channel matrix. Hence, the delay-angular domain channel $\widetilde{\mathbf{X}}_n$ has a low-rank and sparse structure.

\subsection{Joint Activity Detection and Channel Estimation Problem Statement}
We use a signal support $\mathcal{K}$ to denote the collection of active devices at a given time with $K=\left|\mathcal{K}\right|$ denoting the number of active IoT devices. For convenience, we define $\chi_n$ as the activity indicator with ${\chi_n} = 1$ if the $n$th device is active, otherwise, ${\chi_n} = 0$. Towards decreasing the number of samples, the numbers of BS antennas and subcarriers can have low dimensions, i.e. $M_p\leq M$ and $B_p\leq B$. Thus, the received signal $\mathbf{Y}\in \mathbb{C}^{M_p\times B_p}$ at the BS can be expressed as
\begin{eqnarray}\label{recey}
\mathbf{Y}
&=&\sum_{n=1}^{N}\mathbf{P}_M\mathbf{A}_{\theta}\mathbf{X}_n\mathbf{A}_{\tau}^H\mathbf{P}_T\text{diag}({\boldsymbol{\alpha} _n})+\mathbf{Z}\nonumber\\
&=&\sum_{n=1}^{N}\mathbf{B}\mathbf{X}_n\mathbf{A}_n+\mathbf{Z}
=\bar{\mathbf{A}}_\theta\mathbf{X}\bar{\mathbf{A}}_{\tau}^H+\mathbf{Z},
\end{eqnarray}
where $\mathbf{X}_n=\chi_n\widetilde{\mathbf{X}}_n\in \mathbf{C}^{M \times D}$ is the device state matrix of device $n$, $\mathbf{B}=\mathbf{P}_M\mathbf{A}_{\theta}\in \mathbb{C}^{M_p\times M}$, $\mathbf{A}_n=\mathbf{A}_{\tau}^H\mathbf{P}_T\text{diag}({\boldsymbol{\alpha} _n})\in \mathbb{C}^{D\times B_p}$, $\bar{\mathbf{A}}_\tau=[\text{diag}({\boldsymbol{\alpha} _1})^H\mathbf{P}_T^T\mathbf{A}_{\tau},\cdots,\text{diag}({\boldsymbol{\alpha} _N})^H\mathbf{P}_T^T\mathbf{A}_{\tau}]\in \mathbb{C}^{B_p\times DN}$, $\bar{\mathbf{A}}_\theta=\mathbf{P}_M\mathbf{A}_{\theta}\in \mathbb{C}^{M_p\times M}$, and $\mathbf{X}=[\mathbf{X}_0,\mathbf{X}_1,\cdots,\mathbf{X}_{N-1}]\in \mathbb{C}^{M\times DN}$.
$\mathbf{Z}$ denotes an additive white Gaussian noise matrix, $\mathbf{P}_M\in \mathbb{C}^{M_p\times M}$ and $\mathbf{P}_T \in \mathbb{C}^{B\times B_p}$ are sampling matrices which randomly select subset of antennas with cardinality $M_p$ and subset of subcarriers with cardinality $B_p$. The device pilot sequences $\boldsymbol{\alpha}_n\in \mathbb{C}^{B_p}$ are drawn from complex Gaussian distribution and are known to the BS. We assume that all $\boldsymbol{\alpha} _n$ has unit modulus.

Due to a sporadic traffic pattern of the data of IoT devices, $\mathbf{X}$ in (\ref{recey}) is $K$-group sparse. Furthermore, because of the limited scattering nature and small angular spreads of the mmW/THz wideband signal, each group is also sparse. Such a sparsity across groups and within each group is called sparse-group sparsity. In addition, the matrix $\mathbf{X}$ is typically low-rank, namely $\text{rank}(\mathbf{X}) \leq \min\left \{ {M,KL_{\max}} \right \}$ due to low-rank of $\mathbf{X}_n$ and the large BS antenna $M$, where $L_{\max}$ is the maximum value of $L_n$. Then, the simultaneously sparse-group and low-rank signal $\mathbf{X}$ can be recovered by the common approach
\begin{eqnarray}
\label{sgl}
\!\!\!\!&&\!\!\!\!\hat{\mathbf{X}}=\mathop \text{argmin}\limits_{\mathbf{X}}~\nu\left \| \mathbf{X} \right \|_{l_{0}}+\nu_1\text{rank} (\mathbf{X})+\nu_2\sum_{n=1}^{N}\left \| \mathbf{X}_{n} \right \|_{l_F}\nonumber\\
\!\!\!\!&&\!\!\!\! \textrm{s.t.}~~~ \left \| \mathcal{A} (\mathbf{X})-\mathbf{y} \right \|_2\leq \epsilon,
\end{eqnarray}
where the parameter $\nu$, $\nu_1$ and $\nu_2$ are tunable parameters, $l_F$ denotes Frobenius norm of a matrix,  $l_{0}$ and $\text{rank}(\cdot)$ are defined as the number of nonzero elements and the rank of a matrix respectively, $\epsilon$ represents an upper bound on the energy of the noise. The linear mapping $\mathcal{A}: \mathbb{C}^{M\times DN}\rightarrow \mathbb{C}^{B_pM_p} $ obeys $\mathbf{y}=\mathcal{A} (\mathbf{X})+\mathbf{z}$ by straightforward algebra, where $\mathbf{y}$ and $\mathbf{x}$ are obtained by stacking the columns of matrices $\mathbf{Y}$ and $\mathbf{X}$ respectively.

\section{Multi-Rank Aware Sparse JADCE Algorithm}
In this section, we first propose an alternative approximation for problem (\ref{sgl}), then develop a multi-rank aware JADCE algorithm, followed by computational complexity analysis and comparison.

\subsection{Simultaneously Sparse-Group and Low-Rank Approximation}
It is difficult to solve the sparse-group and low-rank problem with an efficient method in (\ref{sgl}) directly due to the combination of the $l_0$ norm, $l_F$ norm, and low rank. To resolve this challenge, we propose to recovery simultaneously sparse-group and low-rank device state matrix by combining $l_0$-norm and low rank under approximate conditions, instead of the method mentioned in (\ref{sgl}).
We start the approximation with the following definition.

{\emph{Definition 1}}: We define a matrix $\mathbf{X}\in \mathbb{C}^{M\times DN}$ to be $u$-sparse-group if
\begin{eqnarray}
\label{jsg}
\left \| \mathbf{X} \right \|_{sg}=\sum_{n \in [\bar{n}]}1_{\left \{ \mathbf{X}_n\neq\mathbf{0} \right \}}p_n^2L_n\leq u,
\end{eqnarray}
holds, where the index set $[DN]$ is partitioned into $\bar{n}$ disjoint sets, and nonzero group $\mathbf{X}_n$ has at most $p_nL_n$ nonzero rows and $p_nL_n$ nonzero columns.

Then, we jointly impose $l_{0}$ norm and low rank on the simultaneously sparse-group and low-rank matrix $\mathbf{X}$, such that
\begin{eqnarray}\label{sgrank}
  \!\!\!\!&&\!\!\!\!\hat{\mathbf{X}}=\mathop \text{argmin}\limits_{\mathbf{X}}\nu\left \| \mathbf{X} \right \|_{l_{0}}+\nu_1\text{rank}(\mathbf{X})\nonumber\\
 \!\!\!\!&&\!\!\!\! \textrm{s.t.}~~~ \left \| \mathcal{A} (\mathbf{X})-\mathbf{y} \right \|_2\leq \epsilon,
\end{eqnarray}
is a viable alternative to the jointly low rank and sparse-group Lasso problem (\ref{sgl}).

Furthermore, we define a specific restricted isometry property (RIP), called sparse-group and low-rank RIP (SG$\&$L-RIP), which is instrumental in building the required theoretical guarantees.

{\emph{Definition 2}}: For positive integers $u$ and $r$, a linear map $\mathcal{A}$ satisfies the SG$\&$L-RIP, if for all simultaneous $u$-sparse-group and rank $r$ matrices $\mathbf{X}$, it is true that
\begin{eqnarray}
(1-\delta_{u,r})\left \| \mathbf{X} \right \|_F\leq \left \| \mathcal{A}(\mathbf{X}) \right \|_2\leq
(1+\delta_{u,r}) \left \| \mathcal{A}(\mathbf{X}) \right \|_2^2,\nonumber
\end{eqnarray}
where $\delta_{u,r}$ is the smallest constant for which the above property holds.

Compared to the traditional definitions of Rank-RIP \cite{rankrip} and Block-RIP \cite{blockrip}, the defined sparse-group and low-rank RIP provides a more restrictive property which holds for the intersection set of the low-rank and the sparse-group matrices. The main theoretical result guaranteeing the SG$\&$L-RIP is summarized in the following theorem.

\emph{Theorem 1}: Let the linear map: $\mathcal{A}: \mathbb{C}^{M\times DN}\rightarrow \mathbb{C}^{B_pM_p} $ obeys the following condition for any $\mathbf{X}\in \mathbb{C}^{M\times DN}$ and $t>1$
\begin{eqnarray}
\label{concen}
P_r(\left \| \mathcal{A}(\mathbf{X}) \right \|_2^2-\left \| \mathbf{X} \right \|_F^2\geq t\left \| \mathbf{X} \right \|_F^2)\leq \text{exp}(-cB_pM_p),
\end{eqnarray}
where $c$ is a fixed parameter for a given $t$. Given integers $u$ and $N$, the map $\mathcal{A}$ satisfies SG$\&$L-RIP of order $\bar{u}:=[1+(t-1)p_{\max}^2L_{\max}]u$ with a constant $\delta_{\bar{u},r}$ with a probability greater than $\geq 1-\bar{C}e^{-\kappa_0}$, if the number of measurements fulfills the following condition:
\begin{eqnarray}
\label{measure}
\!\!\!\!\!\!&&\!\!\!\!\!\!B_pM_p\geq\kappa_1 \left(\Theta \log\frac{N}{\Theta }+\Theta +\Theta p_{\max}L_{\max}\log\frac{D}{p_{\max}L_{\max}}\right.\nonumber \\
\!\!\!\!\!\!&&\!\!\!\!\!\!\left.+\Theta p_{\max}L_{\max}+\left(\Theta p_{\max}L_{\max}+M+1\right)r\right),
\end{eqnarray}
where $\kappa_0$, $\kappa_1$, and $\bar{C}$ are constants for a given $\delta_{\bar{u},r}$, $\Theta =\bar{u}/p_{\min}^2L_{\min}$, $p_{\max}$ and $p_{\min}$ are the maximum and the minimum value of $p_n$ respectively, and $L_{\max}$ and $L_{\min}$ are the maximum and the minimum value of $L_n$ respectively. As a result, the problem (\ref{sgrank}) with a map $\mathcal{A}$ achieves a robust sparse-group and low-rank recovery of order $u$ with a probability greater than $ 1-\bar{C}e^{-\kappa_0}$.

\emph{Remark 1}: It is interesting to find that in the case of a small $ut$ (a reasonable assumption in mmW/THz wideband massive access systems), the required number of measurements derived from Theorem 1 for simultaneously $u$-sparse-group and low-rank matrices are significantly smaller than currently known measurements bounds, which is derived for simultaneously sparse and low-rank matrices \cite{trad2015} by combining $l_{21}$ norm and low rank.

\subsection{Multi-Rank Aware Pursuit}
The JADCE algorithm developed on the set of matrices with low rank compared to the set of all matrices makes the recovery more plausible and efficient, thus it is beneficial to design a rank aware algorithm to solve problem (\ref{sgrank}). However, estimating the rank of $\mathbf{X}$ usually leads to high computational complexity and requires large storage space in practice, due to the large numbers of devices and BS antennas. According to the characteristics of mmW/THz channels analyzed in Section II, the rank of each device state matrix $\mathbf{X}_n$ is not random but equals to the number of paths $L_n$. Notice that the number of paths of mmW/THz channel depends only on the physical propagation properties and the number of paths of mmW/THz channel is usually very limited, e.g. 2-3, which can be measured by channel tracking.

Based on this rank property, problem (\ref{sgrank}) can be transformed as the following problem by exploiting both individual sparse and low-rank structure:
\begin{eqnarray}\label{rankawa}
\!\!\!\!&&\!\!\!\!\mathop \text{argmin}\limits_{\mathbf{X}_n}\frac{1}{2}\left \| \sum_{n=1}^{N}\mathbf{B}\mathbf{X}_n^H\mathbf{A}_n-\mathbf{Y} \right \|_{l_F}^2 +\nu\sum_{n=1}^{N}\left \| \mathbf{X}_n \right \|_{l_1}\nonumber\\
\!\!\!\!&&\!\!\!\!\textrm{s.t.}~~~ \text{rank}(\mathbf{X}_n)=L_{\max},~~~n=1,2,\cdots,N
\end{eqnarray}
Herein, we relax the rank of inactive device state matrices to the maximum number of paths. Since inactive device state matrices infinitely approach to zero matrices, even though we solve a relaxed problem, the rank relaxation would not change the solution to the original problem (\ref{sgrank}). Moreover, due to nonconvex of $l_{0}$ norm, it is quite natural to relax $l_{0}$ norm with $l_{1}$ norm.
Due to multi-rank constraints, problem (\ref{rankawa}) is nonconvex and NP-hard. To tackle this challenge, we exploit the quotient manifold geometry of the product of rank-$L_{\max}$ matrices.

First, we reformulate a problem with Hermitian positive semidefinite variables, instead of directly solving the problem (\ref{rankawa}) with complex asymmetric variables $\mathbf{X}_n$ in complex field, namely
\begin{equation}\label{FAC}
\mathbf{T}_n=\mathbf{S}_n\mathbf{S}_n^H=\left[ \begin{array}{l}
\mathbf{J}_n\mathbf{J}_n^H ~~~\mathbf{J}_n\mathbf{R}_n^H\nonumber\\
\mathbf{R}_n\mathbf{J}_n^H~~~\mathbf{R}_n\mathbf{R}_n^H
\end{array} \right].
\end{equation}
where $\mathbf{S}_n=\left[ \mathbf{J}_n^H,
\mathbf{R}_n^H \right]^H \in \mathbb{C}^{(D+M)\times L_{\max}}$
with full column-rank matrices $\mathbf{J}_n\in \mathbb{C}^{M\times L_{\max}}$ and $\mathbf{R}_n \in \mathbb{C}^{D\times L_{\max}}$. Moreover, the other two auxiliary matrices $
\mathbf{P}_1=\left [ \mathbf{I}_D~~\mathbf{0} \right ]\in\mathbb{C}^{M\times (D+M)}$ and $
 \mathbf{P}_2=\left[ \mathbf{0}~~
\mathbf{I}_{M} \right]^T\in\mathbb{C}^{ (D+M)\times D}
$ are introduced,
where $\mathbf{I}_D$ and $\mathbf{I}_{M}$ denote the identity matrices of order $D$ and $M$, respectively. $\mathbf{X}_n$ satisfies the factorization $\mathbf{X}_n=\mathbf{J}_n\mathbf{R}_n^H=\mathbf{P}_1\mathbf{S}_n\mathbf{S}_n^H\mathbf{P}_2$.

Then, we define the product manifold $\mathcal{M}^N=\mathcal{M}_1 \times \mathcal{M}_2\times\cdots \times\mathcal{M}_N$ as the set of matrices $\left \{ \mathbf{S}_n \right \}_{n=1}^N$, where $\mathcal{M}_n=\left \{ \mathbf{S}_n\in \mathbb{C}^{(D+M)\times L_{\max}}:\text{rank}(\mathbf{S}_n)=L_{\max} \right \}$ is a non-compact stiefel manifold denoting the set of all $(D + M) \times L_{\max}$ matrices whose columns are linearly independent. Consequently, the problem (\ref{rankawa}) is recast as the following unconstrained problem with full column rank optimization variables
$\mathbf{S}_n\in \mathbb{C}^{(D+M)\times L_{\max}}$:
\begin{eqnarray}\label{lift}
\mathop \text{argmin}\limits_{\left \{ \mathbf{S}_n \right \}_{n=1}^N\in \mathcal{M}^N}\!\!\!\!\!&&\!\!\!\!\!f(\left \{ \mathbf{S}_n\right \}_{n=1}^N)=\frac{1}{2}\left \| \sum_{n=1}^{N}\mathbf{B}\mathbf{P}_1\mathbf{S}_n\mathbf{S}_n^H\mathbf{P}_2\mathbf{A}_n-\mathbf{Y} \right \|_F^2 \nonumber\\
\!\!\!\!\!\!\!\!&&\!\!\!\!\!\!\!\!~~~~~~~~~~+\nu\sum_{n=1}^{N}\sum_{i,j}\left(\left |\mathbf{v}_i\mathbf{P}_1\mathbf{S}_n\mathbf{S}_n^H\mathbf{P}_2\mathbf{v}_j  \right |\right.\nonumber \\
\!\!\!\!\!\!\!\!&&\!\!\!\!\!\!\!\!~~~~~~~~~~\left.-\frac{1}{\varrho }
\ln\left(1+\varrho \left |\mathbf{v}_i\mathbf{P}_1\mathbf{S}_n\mathbf{S}_n^H\mathbf{P}_2\mathbf{v}_j  \right |\right)\right),
\end{eqnarray}
where $\varrho >0$ is a tunable parameter, and $\mathbf{v}_i$ and $\mathbf{v}_j$ is used to extract the element located in the $i$-th row and $j$-th column of matrix $\mathbf{P}_1\mathbf{S}_n\mathbf{S}_n^H\mathbf{P}_2$. Notice that $l_1$-norm in problem (\ref{rankawa}) is nonsmooth, which breaks the requirement of a smooth objective function in Riemannian optimization. We therefore have replaced the element of $ \left \| \mathbf{X}_n \right \|_{l_1}$ with the second term in (\ref{lift}), which is the logarithmic smoothing process based on the fact that the function $\left | x \right |-\frac{1}{\varrho}\ln(1+\varrho\left | x \right |)$ is differentiable over $x$ at $0$ \cite{dimen}. After the solution $\hat{\mathbf{S}}_n,\forall n$ of problem (\ref{lift}) is obtained, the original solution $\hat{\mathbf{X}}_n, \forall n$ can be computed by the operation $\hat{\mathbf{X}}_n=\mathbf{P}_1\hat{\mathbf{S}}_n\hat{\mathbf{S}}_n^H\mathbf{P}_2, \forall n$.

To overcome non-uniqueness of the factorization $\mathbf{T}_n=\mathbf{S}_n\mathbf{S}_n^H$ for rank-$L_{\max}$ matrices, we develop a set of  equivalence classes encoding the invariance map with $n=1, 2, \cdots, N$ in an abstract search space in the following form
\begin{eqnarray*}
\!\!\!\!\!\!\!\!&&\!\!\!\!\!\!\!\![\boldsymbol{\mathcal{S}}]=\left \{ [\mathbf{S}_n] \right \}_{n=1}^N\nonumber\\
&=&\left \{ \mathbf{S}_n\mathbf{Q}_n: \mathbf{Q}_n^H\mathbf{Q}_n=\mathbf{Q}_n\mathbf{Q}_n^H=\mathbf{I}, \mathbf{Q}_n\in \mathbb{C}^{L_{\max}\times L_{\max}} \right \}_{n=1}^N.
\end{eqnarray*}
$[\boldsymbol{\mathcal{S}}]$ is also called as the quotient space denoted by $\mathcal{M}^N/\sim$, where a product of non-compact stiefel manifold $\mathcal{M}^N$ is regarded as the
full space. Consequently, if an element $\mathbf{S}_n\in\mathcal{M}_n$ has the matrix characterization
$\mathbf{S}_n\mathbf{Q}_n$, problem (\ref{lift}) can be transformed as
\begin{eqnarray}\label{lift1}
\mathop \text{argmin}\limits_{\left \{ [\mathbf{S}_n] \right \}_{n=1}^N\in \mathcal{M}^N/\sim} f(\left \{ [\mathbf{S}_n]\right \}_{n=1}^N).
\end{eqnarray}
Because the manifold topology of the product manifold is equivalent to the product topology \cite{pa}, the JADCE problem derived from the product manifold $\mathcal{M}^N$ can be processed on individual manifold $\mathcal{M}_n$ and the tangent space to $\mathcal{M}^N$ at $\boldsymbol{\mathcal{S}}$ given by $\mathcal{T}_{\boldsymbol{\mathcal{S}}}\mathcal{M}^N$ can be viewed as the product of the tangent spaces to $\mathcal{M}_n$ at $\mathcal{S}_n$ given by $\mathcal{T}_{\mathcal{S}_n}\mathcal{M}_n
$ for $n=1, 2, \cdots, N$.

In the context of individual manifold, we first give out the Riemannian metric, which is the smoothly varying inner product and invariable along the set $[\mathbf{S}_n]$, namely
\begin{eqnarray}\label{inner}
   g_{\mathbf{S}_n}(\boldsymbol{\xi}_{\mathbf{S}_n},\boldsymbol{\eta}_{\mathbf{S}_n})
&=&\frac{1}{2}\text{Tr}(\boldsymbol{\xi}_{\mathbf{S}_n}^H\boldsymbol{\eta}_{\mathbf{S}_n}+\boldsymbol{\eta}_{\mathbf{S}_n}^H\boldsymbol{\xi}_{\mathbf{S}_n}),
\nonumber \\
&&\boldsymbol{\xi}_{\mathbf{S}_n},\boldsymbol{\eta}_{\mathbf{S}_n} \in \mathcal{T}_{\mathbf{S}_n}{\mathcal{M}_n}
\end{eqnarray}
The individual projection of any direction $\boldsymbol{\xi}_{\mathbf{S}_n}$ onto the horizontal space $\mathcal{H}_{\mathbf{S}_n}$ at $\mathbf{S}_n$ is given by $\Pi_{\mathcal{H}_{\mathbf{S}_n}}(\boldsymbol{\xi}_{\mathbf{S}_n})=\boldsymbol{\xi}_{\mathbf{S}_n}-\mathbf{S}_n\mathcal{B}$,
where $\mathcal{B}$ is a complex matrix of size $L_{\max} \times L_{\max}$, which is the solution of the following Lyapunov equation $
\mathbf{S}_n^H\mathbf{S}_n\mathcal{B}+\mathcal{B}\mathbf{S}_n^H\mathbf{S}_n=\mathbf{S}_n^H\boldsymbol{\xi}_{\mathbf{S}_n}-\boldsymbol{\xi}_{\mathbf{S}_n}^H\mathbf{S}_n$.

According to the previous considerations, it is enough to deduce Riemannian gradient on manifolds represented in the tangent space, which can be expressed as
$
\text{grad}_{\mathbf{S}_n}f=\Pi_{\mathcal{H}_{\mathbf{S}_n}}\left(
\frac{1}{2}\widetilde{\text{grad}}_{\mathbf{S}_n}f\right)
$
where $\widetilde{\text{grad}}_{\mathbf{S}_n}f$ represents the Euclidean gradient of $f(\left \{ \mathbf{S}_n\right \}_{n=1}^N)$ with respect to $\mathbf{S}_n$.

Finally, the conjugate gradient descent approach on the Riemannian space is developed to search the global optimum. Herein, we adopt the truncated spectral initialization, because the dimension of $\mathbf{B}^H\mathbf{Y}\mathbf{A}_n^H$ is high, getting the leading eigenvector of this sample matrix can have large computation. Actually, when $M_pB_p$ is sufficiently large, the leading eigenvector of $\mathbf{B}^H\mathbf{Y}^{tru}\mathbf{A}_n^H$ with norm scaled by $\omega$ can be an approximation of the solution of $(\ref{lift})$. The specific initialization process and the designed multi-rank aware sparse (MRAS) algorithm are detailed in Algorithm 1. Herein, the parameter $o_n^t$ in the Polak-Ribiere form \cite{pa}, $\boldsymbol{\Im}_{\mu^t{\boldsymbol{\eta}_{{\mathbf{S}}_n}^t}}({\boldsymbol{\eta}_{{\mathbf{S}}_n}^t})$
is collinear with $\frac{d}{dt}\boldsymbol{\Im}_{\mu^t{\boldsymbol{\eta}_{{\mathbf{S}}_n}^t}}({\boldsymbol{\eta}_{{\mathbf{S}}_n}^t})|_{\mu^t}={\boldsymbol{\eta}_{{\mathbf{S}}_n}^t}$ and $\mu$ is the step size.

Afterward, we can detect the device activity by defining the activity detector as $\hat{\mathcal{K}}=\left \{ n: \left \| \hat{\mathbf{X}}_n\right \|_F^2\geq v_1\max\limits_{1\leq n \leq N} \left \| \hat{\mathbf{X}}_n\right \|_F^2 \right \}$,
where $v_1=0.1$ denotes the ratio of the minimum and maximum amplitudes of the channel coefficients.

\begin{algorithm}[h]
\caption{Multi-Rank Aware Sparse Recovery for JADCE.}
\label{alg3}
\begin{algorithmic}[1]
\STATE \textbf{Input}: Received signal $\mathbf{Y}$, matrices $\mathbf{B}$, $\{\mathbf{A}_n\}_{n=1}^N$ and step size $\mu$,the device index $n=N$, and the iteration index $t=T$.  \\
\STATE \textbf{Truncated Initialization Evaluation}: \\
\STATE Set the element of $\mathbf{Y}^{tru}$ as
$y^{tru}_{l,m}=\begin{cases}
&y_{l,m}, \text{ if } y_{l,m}\leq \frac{\omega}{M_pB_p}\sum_{l=1}^{M_p}\sum_{m=1}^{B_p}y_{l,m} \\
&0,~~~ \text{ if } \text{otherwise}
\end{cases}$,
where $y_{l,m}$ denotes the $(l,m)$th element of $\mathbf{Y}$.
\WHILE{$n\geq 1$}
\STATE Let $\widetilde{\mathbf{J}}_n^0\boldsymbol{\Sigma}_ n^0\widetilde{\mathbf{R}}_n^0$ to be the rank-$L$ eigendecomposition of $\mathbf{B}^H\mathbf{Y}^{tru}\mathbf{A}_n^H$. Set $\mathbf{S}_n^0=\left[ (\mathbf{J}_n^{0})^H,
(\mathbf{R}_n^{0})^H \right]^{H}$ with $\mathbf{J}_n^0=\widetilde{\mathbf{J}}_n^0\sqrt{\boldsymbol{\Sigma}_ n^0}$ and $\mathbf{R}_n^0=\sqrt{\boldsymbol{\Sigma}_ n^0}\widetilde{\mathbf{R}}_n^0$.
\STATE Update $n \leftarrow n-1$.
\ENDWHILE
\WHILE{$t\geq 1$}
\STATE
$\forall n$: $\mathbf{S}_n^{t+1}=\mathbf{S}_n^{t}+\mu\left(-\text{grad}_{\mathbf{S}_n^{t}}f+o_n^t\boldsymbol{\Im}_{\mu^t{\boldsymbol{\eta}_{{\mathbf{S}}_n}^t}}({\boldsymbol{\eta}_{{\mathbf{S}}_n}^t})\right)$,
\STATE Update $t \leftarrow t-1$.
\ENDWHILE
\STATE \textbf{Output}: $\forall n$: $\hat{\mathbf{S}}_n=\mathbf{S}_n^{t+1}$, $\hat{\mathbf{X}}_n=\mathbf{P}_1\hat{\mathbf{S}}_n\hat{\mathbf{S}}_n^H\mathbf{P}_2$, $\hat{\mathbf{H}}_n=\mathbf{A}_{\theta}\hat{\mathbf{X}}_n\mathbf{A}_{\tau}^H$.
\end{algorithmic}
\end{algorithm}

In what follows, we analyze the computational complexity of the proposed algorithm. The computational burden of the proposed MRAS algorithm mainly has two
parts when $D$ and $M$ have the same order: 1) the computational complexity of the $\text{grad}_{\mathbf{S}_n}f$ has two cases, i.e. when $M<B_p$, it is $\mathcal{O}(M_pB_pDN)$, otherwise, the complexity is $\mathcal{O}(M_pDMN+B_pDMN)$. 2) the computational complexity of Riemannian metric in (\ref{inner}) is $\mathcal{O}(L^2D)$.

In this paper, we compare the proposed MRAS algorithm with four baseline algorithms from the computational complexity aspect, including AMP algorithm \cite{amp} which leverages large-scale fading coefficients and the statistics of the wireless channel to improve the detection performance, fast iterative shrinkage-thresholding algorithm (FISTA) \cite{fista} which is a classical optimization algorithm to minimize convex functions, OMP algorithm which is a greedy algorithm proposed in \cite{OMP}, and the baseline in (\ref{sgl}) which can be solved by replacing $\text{rank}(\mathbf{X})$ by the nuclear norm and then reformulating as a semidefinite programming problem. The comparison results are shown in Table 1. It can be seen that the complexity scaling of the proposed MRAS algorithm is superior to the four baseline algorithms, implying lower complexity in the high dimensional regime.

\begin{table}[h]
\centering
\caption{The Computation Complexity Comparison of Considered Schemes.}
\label{tab1}
\begin{tabular}{cc}
\hline
Schemes     & Computational Complexity   \\
\hline
Proposed MRAS & $\begin{cases}
 \mathcal{O}(M_pB_pDN),& \text{ if } M<B_p  \\
 \mathcal{O}(M_pDMN+B_pDMN),& \text{ if } M\geq B_p
\end{cases}$
  \\
AMP \cite{amp} & $\mathcal{O}(B_pM_pNDM)$  \\
FISTA \cite{fista} &  $\mathcal{O}(B_pM_pNDM)$  \\
OMP \cite{OMP}&  $\mathcal{O}(N^3L^3p^6+B_pM_pNDM)$  \\
Baseline in (\ref{sgl})&  $\mathcal{O}(B_p^6M_p^6+N^6D^6M^6)$  \\
\hline
\end{tabular}
\end{table}

\section{Numerical Results}
In this section, we investigate the performance of the proposed MRAS algorithm in terms of activity error rate (AER) and normalized mean squared error (NMSE). The AER includes miss detection probability defined as the probability that an active device is detected as inactivity, and the false-alarm probability defined as the probability that an inactive device is detected as activity. The NMSE is calculated as $\frac{\sqrt{\sum _{n=1}^N\left \|\mathbf{H}_n- \hat{\mathbf{H}}_n \right \|_F^2}}{\sqrt{\sum _{n=1}^N\left \| \hat{\mathbf{H}}_n \right \|_F^2}}$. We consider a simulation scenario where the BS employs a ULA antenna array with $M=64$, the total number of subcarriers is set to $B=1300$, the channel delay is set to $D=64$. All $N=20$ devices are randomly distributed in the service area and among which $K=6$ devices are active. Parameters $\varrho$ and $\nu$ are set to $1/0.039$ and $0.3$ respectively. The maximum number of paths is set to $L_{\max}=2$ and  SNR set as 25 dB.

\begin{figure}[h]
  \centering
\includegraphics [width=0.37\textwidth] {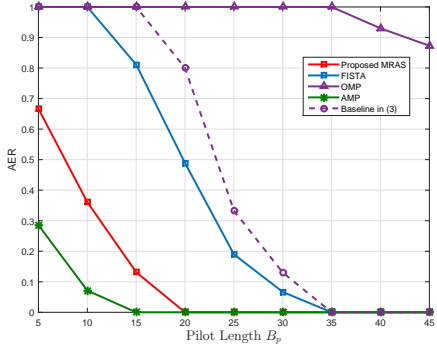}
\caption{The activity error rate for different length of pilot.}
\label{AERpilot}
\end{figure}

We first compare the performance of the proposed MRAS algorithm and the above four state-of-the-art algorithms. Fig. \ref{AERpilot} plots the AER performance comparison against the pilot length $B_p$ . It is observed that increasing the length of pilot sequences substantially decreases the error probability for all detection algorithms. Yet, for the proposed MRAS algorithm, the gain by increasing $B_p$ is obvious. For instance, when the number of subcarriers is greater than 20, the activity error rate of the proposed MRAS reaches 0, which outperforms the OMP algorithm, the FISTA algorithm and the baseline in (\ref{sgl}). This is because the proposed MRAS algorithm well incorporates the joint sparse and low rank characteristics of mmW/THz channel for efficiently decreasing the search space of the JADCE problem. Compared to the AMP algorithm, the proposed MRAS algorithm performs worse but has lower complexity. Moreover, the AMP algorithm needs the prior sparsity information and statistics of the channel vectors, which are difficult to be obtained in practice due to the sporadic traffic or spatial correlation of the channel.

\begin{figure}[h]
  \centering
\includegraphics [width=0.37\textwidth] {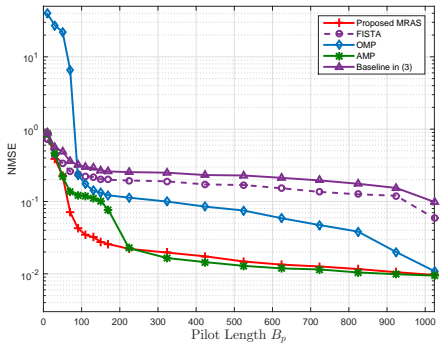}
\caption{The NMSE for different length of pilot.}
\label{pilotnmse}
\end{figure}

In Fig. \ref{pilotnmse}, we present the NMSE performance comparison against the number of selected subcarriers $B_p$. We note that the curve of the proposed algorithm for the NMSE decreases rapidly when $B_p$ exceeds a certain threshold. The NMSE of the proposed MRAS algorithm achieves a high channel estimation accuracy, which is superiority over OMP, FISTA and the baseline in (\ref{sgl}).
The NMSE gap between the proposed algorithm and the AMP algorithm narrows with the increase of $B_p$. This demonstrates that the proposed algorithm can obtain reasonable channel estimation accuracy with relatively short pilot sequences.

\begin{figure}[h]
  \centering
\includegraphics [width=0.37\textwidth] {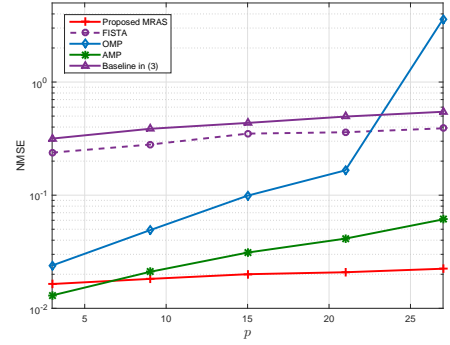}
\caption{The NMSE for different delay and angular spreads.}
\label{snrnmse}
\end{figure}

Fig. \ref{snrnmse} displays the NMSE versus different spread $p$. It is seen that the proposed algorithm by properly exploiting multiple low rank and sparse can achieve the optimal NMSE performance in large spread regions. However, the other four algorithms degrade seriously. Since the mmW/THz channel usually has a large spread due to high angular and delay resolutions, the proposed MRAS algorithms is appealing in wideband massive access.

\section{Conclusion}
This paper has studied a grant-free random access scheme for mmW/THz wideband cellular IoT networks with sporadically active devices. The low-rank and sparse characteristics in delay-angular domain of mmW/THz channels were investigated and then explored to design a JADCE algorithm for wideband massive access. Theoretical analysis proved that the proposed algorithm can shorten the required length of pilot sequences and has lower computational complexity compared to baseline algorithms. Simulation results shew that the proposed algorithm can almost achieve the near-optimal detection and estimation performance.

\end{document}